# AN AXISYMMETRIC DISTRIBUTION FUNCTION
# FOR THE GALACTIC BULGE

## Konrad Kuijken[1,2]

Harvard–Smithsonian Center for Astrophysics, 60 Garden St., Cambridge, MA 02138

and

Visiting Scientist, Dept. of Theoretical Physics, University of the Basque Country, Bilbao, Spain

March 17, 1994

## ABSTRACT

We describe a method for parameterizing two-integral distribution functions, based on triangular tesselations of the integral plane. We apply the method to the axisymmetric isotropic rotator model for the Galactic bulge of Kent (1992), and compare the results with observations of proper motions in Baade's Window, and with radial velocity surveys. In spite of mounting evidence from surface photometry and from study of the gas kinematics that the Galactic bulge is not axisymmetric, the stellar kinematics in Baade's Window are very similar to those of an isotropic oblate rotator. Another field at large radius does not fit this model, though. In any case, the edge-on kinematics of a hot stellar population are a poor handle on the existence or otherwise of a bar.

## 1   Introduction

In the light of mounting evidence that the Galactic bulge is rather barred, (Blitz and Spergel 1991, Binney et al. 1991, Whitelock, Feast & Catchpole 1991, Weinberg 1992), it seems opportune to check to what extent the stellar kinematics of the bulge stars reflect this distortion. De Zeeuw (1993) recently addressed this question, and concluded that, at least as far as the first two moments of the velocity distribution are concerned, simple axisymmetric analytic models such as the one by Kent (1992) can reproduce the observed kinematics of stars seen in Baade's Window and other central fields reasonably well. This would suggest that either the indications for bulge non-axisymmetry are misleading, or that triaxial and axisymmetric models for the bulge predict quite similar stellar kinematics in these fields. In the absence of triaxial models for the bulge, we explore the axisymmetric models a little further in this paper. In particular, we will construct in as much detail as possible the full phase space distribution function (DF) for the bulge stars, assuming that the bulge can be well-modelled as an oblate isotropic rotator (see §2).

---

[1]Hubble Fellow

[2]I: kuijken@cfa.harvard.edu

Oblate isotropic rotator models are often used in astronomy (e.g., Binney, Davies & Illingworth 1990), because they offer a convenient way of closing the infinite hierarchy of Jeans equations that relate the various velocity moments throughout a galaxy (Binney and Tremaine 1987). Such models have the nice feature that the first few moments of the velocity distribution everywhere follow straightforwardly from the Jeans equations, and can be calculated without knowledge of the DF. It is therefore relatively simple to compare observed velocity moments with the predictions of the oblate isotropic rotator. But in fact the full DF is determined in these models. In principle, therefore, knowledge of the DF would allow the comparison with observed kinematics to be extended beyond low-order velocity moments. It would also make it possible to test for non-negativity of the DF, a physical requirement that is by no means guaranteed to be satisfied. Furthermore, the fact that the shape of the Doppler broadening functions in galaxies can now be measured with a variety of techniques (e.g., Kuijken & Merrifield 1993 and references therein) motivates the calculation of similar quantities from models.

The analytic calculation of the DF for oblate isotropic rotators is not generally possible: a complicated integral equation relating the density in the meridional plane to the distribution function, written in terms of stars' specific energy and angular momentum, needs to be inverted. Analytic work is progressing, though: after the original formulation of the inversion in terms of double inverse Lapace transforms by Lynden-Bell (1962), Hunter (1975) recast the problem in terms of an, in some circumstances more manageable, inverse Stieltjes transform, and recently Hunter and Qian (1993) have rewritten the inversion as a direct contour integral in the complex plane. Unfortunately, in spite of these results, none of the analytic methods are applicable to general cases, since all require analytic continuation of the density into the complex plane.

The limitations of the analytic methods prompted us to explore numerical techniques for construction of the DF. In spite of the greater computational cost, this strategy has the advantage that there are no restrictions on the shapes of the density profiles, so that a model approximating observations as closely as possible can be constructed. We will use a numerical method to study the DF that corresponds to a model very similar to the one Kent (1992) constructed for the Milky Way bulge based on a $2.4\mu m$ Spacelab map of the Galactic plane (Kent et al. 1992)

We summarize the oblate isotropic rotator model, and the integral equation that needs to be solved in order to recover the DF, in §2. The numerical inversion technique is set out in §3, and then applied to construct a kinematic model for the central few kpc of the Milky Way in §4. Comparisons of the model kinematics with radial velocity and proper motion data are also made there. Finally, the implications of the result for the structure of the bulge are discussed in §5.

A special case of the integral inversion, which applies on the minor axis of any axisymmetric system (whether the DF is two-integral or not) is discussed in the Appendix.

## 2    THE TWO-INTEGRAL DISTRIBUTION FUNCTION FOR OBLATE ISOTROPIC ROTATORS

There are generally many DFs which have the same mass density in a given gravitational potential (Schwartzschild 1979), and unless the potential is of a special form most of these cannot be written down analytically. The reason for the many different possible DFs is that most realistic three-dimensional potentials support three independent *isolating integrals of motion* $I_i$ (i.e., single-valued functions of the phase space variables which are conserved along stellar orbits), and therefore the DF $f(R, z, v_R, v_\phi, v_z)$ (in cylindrical polar coordinates aligned with the symmetry axis of the system) can be written as a function of the isolating integrals only, $f(I_1, I_2, I_3)$. In an axisymmetric system all three isolating integrals are axisymmetric, and therefore the requirement that the density corresponding to the DF be the given density $\rho(R, z)$ is a two-dimensional constraint on a three-dimensional function $f(I_1, I_2, I_3)$. The inverse problem of finding a DF for a given density figure can only be made well-determined by reducing the dimensionality of the allowed DFs.

In axisymmetric systems, two of the integrals of motion are direct consequences of the symmetries of the Hamiltonian: they are the specific energy, $E = \Psi(R, z) + \frac{1}{2}v_R^2 + \frac{1}{2}v_\phi^2 + \frac{1}{2}v_z^2$, and the $z$-component of the specific angular momentum, $L = Rv_\phi$. Because the third integral is usually not an easily accessible function (it may not even exist through all of phase space), most analytic studies of axisymmetric systems concentrate on DFs depending only on the two classical integrals, $f(E, L)$. This choice also accomplishes the dimensionality reduction required if the inversion problem outlined above is to be well-posed. In this paper we will restrict ourselves to the two-integral models.

Many investigations center on the Jeans equations, which relate the first and second velocity moments and their gradients to the potential and density in the system (see e.g., Binney & Tremaine 1987). Two-integral models have the special properties that the $R$- and $z$-velocity dispersions everywhere are the same, and the value of this dispersion is uniquely determined by the density and gravitational potential of the model. Moreover, the dispersion is simply calculated using the Jeans equations. If the system is flattened, a suitable amount of rotation about the symmetry axis can be introduced, such that the $\phi$-velocity dispersion is everywhere equal to the dispersion in the other two components. This leads to the so-called *isotropic oblate rotator* models. Such models are often used in analyses of observed stellar systems (e.g., Binney, Davies & Illingworth 1990, van der Marel 1991). They provide a useful foil for the interpretation of observed kinematics, and often fit the data surprisingly well, in spite of their simplicity and gross non-uniqueness. Nevertheless, detailed analysis of well-observed edge-on elliptical galaxies by Merrifield (1991) reveals that the central assumption of a two-integral DF is not always justified.

In fact, the two-integral assumption has much stronger consequences then the specification of the mean azimuthal velocity and radial velocity dispersion everywhere: as shown by Lynden-Bell (1962), in such models the entire even part of the DF, $f_+ \equiv f(E, L) + f(E, -L)$, is determined by the density and potential. The further assumption of isotropicity, equivalent to specifying a particular mean azimuthal velocity everywhere, fixes the odd part of the DF, $f_- \equiv f(E, L) - f(E, -L)$. Therefore isotropic oblate rotators have a uniquely defined DF. The goal of this paper is to present a numerical method for calculating the DF for this class of models for general densities and potentials, and to apply it to the oblate isotropic rotator model that was constructed for the Milky Way bulge by Kent (1992).

### 2.1   The basic integral equations

In the $(E, L)$ plane, the physically accessible orbits lie in the wedge-shaped area above the curve $E = E_{circ}(L)$, where $E_{circ}(L)$ is the energy of a circular orbit in the equatorial plane of angular momentum $L$ (see figure 2 below).[1]

The density $\rho(R, z)$ corresponding to the distribution function $f(E, L)$ is

$$\rho(R, z) = \int dv_R dv_\phi dv_z f(E, L) = 2\pi R^{-1} \int_{-\infty}^{\infty} dL \int_{\Psi(R,z)+\frac{1}{2}\frac{L^2}{R^2}} dE \, f(E, L) \qquad (1)$$

when $R \neq 0$. At $R = 0$,

$$\rho(0, z) = 2^{\frac{5}{2}} \pi \int_{\Psi(0,z)}^{\infty} dE \, [E - \Psi(0, z)]^{\frac{1}{2}} f(E, 0). \qquad (2)$$

In typical applications the density is related directly to the potential $\Psi$ by Poisson's equation, but other sources of gravity (such as a dark halo) could also be included. In any case, if the potential is known, a given DF $f(E, L)$ determines the density completely.

The inversion of eqs. (2), starting from the observed density and resulting in the distribution function $f$, can be written formally as a double inverse Laplace transform (Lynden-Bell 1962), in terms of an inverse Stieltjes transform (Hunter 1975), or as a contour integral in the complex plane (Hunter and Qian 1993). Unfortunately, none of these methods are applicable to general cases.

### 2.2   The isotropic rotator

Adding any term which is odd in $L$ to the DF $f(E, L)$ does not change the density in Eq. (1). The inversion problem therefore does not have a unique solution, (the physical reason is that reversing the motion along any of the orbits changes $L$ but does not affect the density). By also specifying the mean azimuthal velocity, which is related to the DF by

$$\rho \overline{v}_\phi(R, z) = 2\pi R^{-2} \int_{-\infty}^{\infty} L \, dL \int_{\Psi(R,z)+\frac{1}{2}\frac{L^2}{R^2}} dE \, f(E, L) \qquad (3)$$

for $R \neq 0$ ($\overline{v}_\phi = 0$ at $R = 0$), the odd part of the DF is also fixed. There are various choices that can be made: $\overline{v}_\phi$ can be set to zero, for instance, or to the 'maximum streaming' value that obtains if all orbits have positive angular momentum. An intermediate case is the so-called 'isotropic rotator' model. This model is the one in which the r.m.s. dispersion of the azimuthal velocity about the mean streaming matches the dispersion in the $R$- and $z$-directions[2]. The mean streaming of an isotropic rotator model is simple to evaluate using the Jeans equations (Binney and Tremaine 1987), with the result

$$\overline{v}_\phi^2 = \frac{R}{\rho}\frac{\partial(\rho\sigma^2)}{\partial R} + R\frac{\partial\Psi}{\partial R}, \qquad (4)$$

where

$$\sigma^2 = \frac{1}{\rho}\int_z^\infty \rho(R,z')\left.\frac{\partial\Psi}{\partial z}\right|_{(R,z')} dz'. \qquad (5)$$

is the (isotropic) velocity dispersion of each velocity component.

---

[1] at least if the potential increases as a function of $R$ and $|z|$, as is almost always true in a galactic context.

Because the velocity dispersion and mean streaming velocity are so simple to calculate once the potential and density are given, isotropic rotator models have been used a great deal in astronomy. However, it has not often been emphasized that in fact the assumptions that lead to this model fully determine the entire distribution function (and therefore the distribution of the three velocity components at every point in space), and not just the first two velocity moments. Kinematic data can therefore be compared with such a model in great detail.

An illustration of the possible pitfalls that can be encountered when the Jeans equations are used without consideration of the DF is provided by prolate Binney potentials, of the form

$$\Psi = \tfrac{1}{2}v_c^2\ln(R_c^2 + R^2 + (z/q)^2) \qquad (6)$$

with $q > 1$. Evans (1993a) constructed the analytic two-integral DF for these models, and showed that prolate models only exist (i.e., have non-negative DF) in the narrow interval $q < 1.08$. Nevertheless, the density in prolate models corresponding to the potential (6) is everywhere positive, and so is the velocity dispersion derived from the Jeans equations (eq. 5). Thus a 'blind' use of the dispersions returned by the Jeans equation (for example, comparing them to gaussian velocity dispersions derived from fitting absorption-line spectra) could return essentially meaningless results.

A different caution applies to the practice of modelling real stellar systems as isotropic oblate rotators: such models are, in fact, very restrictive. Both ignoring the third integral, which is almost invariably present in realistic potentials for stellar systems, and dividing the prograde and retrograde orbits' populations just so that the azimuthal dispersion matches the radial and vertical ones represents a very special choice of DF. For instance, the two-integral assumption means that the number of stars on planar, excentric orbits is coupled with the number of stars on radially thin but vertically extended 'shell' orbits, and in very flattened models isotropy of the velocity dispersion requires that the stars are divided into two distinct counter-rotating components (a fact that would not be evident from the Jeans equations alone). There is nothing wrong *per se* with these choices, but they are very restrictive.

---

[2]This statement about the second velocity moments does not imply that the velocity *distribution* is isotropic about the mean streaming.

From the point of view of combining Galactic 'components', such as bulge, disk, halo, etc., the isotropic rotator models are also rather artificial, since the linear superposition of two isotropic rotators with different mean streaming does not result in a new isotropic rotator. Thus, building an isotropic rotator model for a bulge+disk system, for example, amounts to assuming that each of the components is anisotropic in just such a way that the combination has an isotropic velocity distribution.

In spite of the caveats, it is advantageous to study the properties of two-integral DFs in detail. The assumption of an isotropic rotator model does, after all, specify a *unique* DF (up to the sign of its spin), and comparison of this model with, for example, projected radial velocities or proper motions can be made in more detail than what can be learned from low-order velocity moments. Numerical techniques for contructing two-integral DFs should also eventually be extended to the more realistic three-integral models. In any event, it is always worthwhile to calculate the DF itself, since this is the only way to verify that it is non-negative: an analysis through moment equations alone in no way guarantees non-negativity.

## 3   Recovering the Distribution Function Numerically

Unfortunately, at the moment there is no generally applicable analytic method for the construction of two-integral models, though progress is being made in that area (Lynden-Bell 1962, Hunter 1975, Hunter and Qian 1993). These authors list several analytic models, and Evans (1993a, 1994) has recently used Lynden-Bell's method to construct the DF for Binney's logarithmic potential with a core as well as more general potentials stratified on ellipsoids. We could approximately invert the bulge data by fitting one of these models to the density figure of the bulge — but unfortunately none of them are particularly good descriptions of the density and potential in the central few kiloparsecs of the Galaxy. The best analytic model that has been fitted to the bulge so far is due to Evans (1993b), but it is not a perfect match. We therefore use instead a more numerical inversion of eqs. (1) and (3). Though quite computationally intensive, this inversion can be applied with equal ease to complex density-potential pairs, including those which cannot be written down analytically.

Since the inversion is somewhat unstable, numerical methods which impose some regularity or smoothness condition on the inverted DF are essential if they are to be of practical use. In this attempt, we employ a parameterization of the DF in continuous bilinear segments. This is the lowest-order continuous parameterization, and can be considered as the next refinement beyond a 'histogram' parameterization, in which the DF is taken constant within different cells of the integral plane. Essentially the same method was used by Merrifield and Kuijken (1994) for the description of a thin disk DF. Different approaches to describing smooth DFs are described by Merritt (1993), and by Dejonghe (1993).

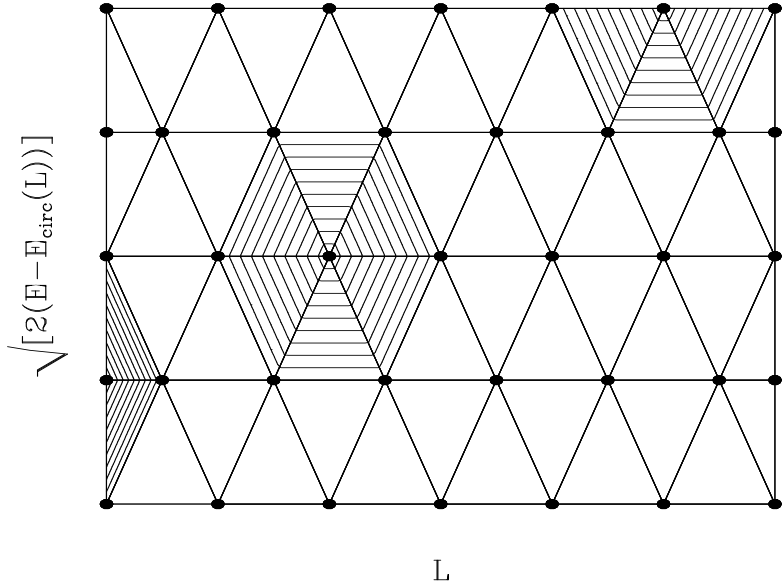

FIG. 1.–An example of a bilinear-tesselation grid. In addition, linearly spaced contours of three elementary components are shown.

### 3.1 Bilinear Tesselation with Quadratic Programming

A bilinear tesselation is simply described as follows (Merrifield & Kuijken 1994). First, a grid of triangles is laid down on the integral plane (e.g., Fig. 1). The DF will be approximated by bilinear functions $f = a + bx + cy$ inside each triangular cell, chosen so that they match continuously at cell boundaries. Since a bilinear function requires three coefficients, $f(x, y)$ is defined uniquely by bilinear interpolation between the three vertices of whatever grid triangle $(x, y)$ happens to lie in (this explains the need for a triangular, rather than a square, grid). Thus, the grid and the values of $f$ on the grid point determine the bilinear-tesselation DF completely. $f$ can be viewed as the sum of elementary basis functions, each of which is non-zero at all grid points but one, and linearly interpolated in between; our task is then to find the amplitudes for each basis function. Clearly the 'resolution' (i.e., the size of the smallest features that can be described with this description of the DF) is set by the density of the grid.

The physical boundary of the $(E, L)$-plane is inconveniently curved (see figure 2 below). Because it is most convenient to impose a grid with a rectangular boundary, it is profitable to transform to coordinates in which this curve is straightened out. This is accomplished by the transformation

$$E \to \mathcal{E}_{\mathrm{norm}} \equiv \frac{E - E_{\mathrm{circ}}(L)}{E_{\mathrm{cut}} - E_{\mathrm{circ}}}. \tag{7}$$

At every angular momentum, this has the effect of rescaling the energy to a variable, $\mathcal{E}_{\mathrm{norm}}$, which runs from zero at the circular energy to unity at a cutoff energy $E_{\mathrm{cut}}$. In order to increase the resolution near the circular orbits, a transformation $E \to \mathcal{E}_{\mathrm{norm}}^n$ with $0 < n < 1$ could be used instead; however, the correspondingly lower resolution that results at higher energies can be problematic, and generally such a transformation is no substitute for using a sufficiently fine grid.

A 'solution' to the inverse problem for the DF is then an optimization of the values $f_\alpha$ of the DF at the grid points $\alpha$ (since the $f_\alpha$ determine the DF everywhere). A physical solution must involve non-negative $f_\alpha$ only — clearly the interpolated values of the DF between the grid points will also be non-negative then. The function to be optimized can take many forms: typically it is a sum of square deviations between some measured data and the corresponding model prediction. Measurements could be observed brightnesses, distributions of radial velocities, or even galactic absorption-line spectra. In our case, we fit the model DF to the density and to the mean streaming velocity corresponding to an isotropic rotator model, given by eqs. (1) and (3,4). For stability, these 'data' values are calculated on a grid $(R_i, z_i)$ containing many more points $i$ than the $(E_\alpha, L_\alpha)$-grid used to describe the DF.

A least-squares fit to the desired $\rho_i$ and $\rho\overline{v}_{\phi i}$ on the grid involves minimizing the expression

$$\chi^2 = \sum_i w_i[\rho_i - \rho(R_i, z_i; f_1 \ldots f_N)]^2 + \sum_i w_i[\rho\overline{v}_{\phi i} - \rho\overline{v}_\phi(R_i, z_i; f_1 \ldots f_N)]^2. \qquad (8)$$

Since both $\rho(R_i, z_i)$ and $\rho\overline{v}_\phi(R_i, z_i)$ depend linearly on the DF, and hence are also linear functions of the $f_\alpha$, $\chi^2$ is a quadratic function of the $f_\alpha$. Since we wish to minimize $\chi^2$ subject to the conditions that the $f_\alpha$ are non-negative (since they represent phase space densities), the optimal $f_\alpha$ can be found as the solution of a Quadratic Programming (QP) problem, i.e., the minimization of a quadratic function subject to a number of linear constraints (e.g., Dejonghe 1989). Efficient algorithms for obtaining this solution (which is generally unique) are implemented in the NAG and IMSL libraries.

The use of these basis functions offers some advantages over other possible choices, such as Fricke components (e.g., Dejonghe 1993). Non-negativity is much easier to ensure (with more extended, overlapping components such as polynomials in $E$ and $L$, positivity of the DF has to be checked on a fine grid since some of the components may well have negative coefficients), and the situation in which very good data in one part of phase space dictates the fit in other regions is avoided by having compact phase space basis functions. Thus only the *physical* correlations between the properties of the system in different parts of phase space enter into the fit.

In summary, the combination of a parameterization in bilinear segments and the QP algorithm provides a general approach for constructing the unique two-integral distribution function corresponding to a choice of mean streaming model. The inherent instabilities in the inversion of the fundamental integral equations (1) and (3) are suppressed in this method by the finite resolution of the grid on which the DF is defined, and the requirement that the phase space density not be negative anywhere on the grid.

### 3.2   The minor axis

We saw earlier (eq. 2) that the density on the minor axis $\rho(0, z)$ is related to the phase space density at zero angular momentum, $f(E, 0)$. This equation is of Abel form, and can be inverted to yield

$$f(E, 0) = -\frac{1}{2\pi^2}\frac{d}{dE}\int_E^\infty d\Psi\, \frac{d\rho(0, z)}{d\Psi}[2(\Psi - E)]^{-\frac{1}{2}}. \tag{9}$$

(This is also Eddington's formula, which relates the density and the DF in spherical isotropic systems). Thus, in any two-integral model (it need not be the isotropic rotator) the phase space density of plunging orbits is everywhere determined by the minor axis density (and the potential).[3] Because it relies critically on the lack of a third isolating integral, this relation is of limited practical importance in an observational context; nevertheless it is a very quick first check whether a particular density-potential pair can be described by a non-negative two-integral model. A sufficient, but not quite necessary, condition is clearly $d^2\rho/d\Psi^2 \geq 0$.[4]

## 4   KENT'S ISOTROPIC ROTATOR MODEL FOR THE GALACTIC BULGE

Using a map of the infrared luminosity of the galactic bulge region made with the Spacelab Infrared Telescope (Kent et al. 1992), Kent, Dame and Fazio (1991) constructed a model for the stellar emission of the central few kiloparsecs of the Galaxy. They found that the data could be modelled well with a two-component model, consisting of a 'bulge' and a 'disk.' The disk's $2.4\mu$m emissivity, $\nu_D$, was modelled as a double exponential,

$$\nu_D(R, z) = 3.0\exp[-R/(3.001\text{kpc})]\exp[-|z|/(0.165\text{kpc})]\mathcal{L}_\odot\text{pc}^{-3} \tag{10}$$

and the bulge was described as a flattened object with boxy isophotes whose emissivity $\nu_B$ is constant on surfaces of constant $s^4 = R^4 + (z/0.61)^4$:

$$\nu_B(R, z) = \begin{cases} 3.53K_0[s/(0.667\text{kpc})]\mathcal{L}_\odot\text{pc}^{-3} & \text{if } s > 0.938\text{kpc}, \\[2mm] 1.04\times10^6[s/(0.482\text{kpc})]^{-1.85}\mathcal{L}_\odot\text{pc}^{-3} & \text{if } s < 0.938\text{kpc}. \end{cases} \tag{11}$$

Kent (1992) then calculated the velocity dispersions and mean streaming for the isotropic rotator model corresponding to different choices of disk and bulge mass-to-light ratios, using eqs. (4) and (5). He found that an infrared bulge mass-to-light ratio of about 1 produced a model which was in excellent agreement with stellar velocity dispersions measured towards the central few kiloparsecs of the Galaxy.

---

[3] This is no longer so in three-integral models, which can be anisotropic even on the $z$-axis.

[4] It is simple to show that on the $z$-axis the integrated $z$-velocity distribution $\int f dv_R dv_\phi$ satisfies the one-dimensional CBE, and therefore depends only on the function $\rho(\Psi)$. Eq. (9) then follows as the (Abel) deprojection of the isotropic velocity distribution into the joint distribution of the three components. See also the Appendix.

TABLE 1.-Grid parameters

| | |
|---|---|
| $L$-grid limits | $-210$ to $210\,\mathrm{km\,s^{-1}kpc}$ |
| Number of $L$-points | 60 |
| $\mathcal{E}_{\mathrm{norm}}$-grid limits | 0 to 1 |
| Number of $\mathcal{E}_{\mathrm{norm}}$-points | 15 |
| Cutoff energy | $-3.7 \cdot 10^4 (\mathrm{km\,s^{-1}})^2$ |

TABLE 1.—Grid parameters for the bilinear-segment reconstruction of the distribution function of Kent's (1992) bulge model. The angular momentum limit corresponds to a circular orbit of radius 1.2 kpc.

We now construct the stellar distribution function that corresponds to Kent's model. As explained in §2, this DF is fully specified by the assumptions that are made in the calculation of the isotropic rotator velocity dispersions.

We make three small changes to Kent's model. The first is to ignore his central black hole of $3 \times 10^6 \mathcal{M}_\odot$, since we will not be able to resolve the central kinematics very well in any case. As Kent shows, the kinematics are not affected by this central mass outside $R = 10\mathrm{pc}$. Instead, we soften the bulge component by replacing $s$ by $(s^2 + 0.001\mathrm{kpc}^2)^{1/2}$ in eq. (11). The second change is that we calculate the potential due to the actual mass distribution of the bulge (eq. 11), rather than using an ellipsoidal approximation to it, as Kent does. The third change avoids the logarithmic singularity that arises in the distribution function for one-dimensional self-gravitating exponential disks. To achieve this, we replace the exponential $z$-dependence of the disk density in eq. (10) by a factor of $\frac{1}{2}\mathrm{sech}\,z$. This function has been found by van der Kruit (1988) to be a good description of many edge-on galactic disks. The difference between the exponential and sech functions is most significant in the lowest one or two scale heights, where the effects of extinction are at any rate still uncertain. For the same infrared flux emissivity at large altitude, the sech model has a disk surface density which is a factor $\frac{\pi}{4} \simeq 0.79$ lower than that of the exponential model.

The first step in the calculation of a two-integral model for a given axisymmetric density $\rho(R, z)$ is to calculate the gravitational potential. The potentials for the bulge and disk components were obtained by first fitting axisymmetric spherical harmonics to the density on spherical shells, and then calculating the potential harmonic by harmonic (see Binney and Tremaine 1987, eq. 2-208). This treatment is more general, but also numerically more involved, than the fairly common practice of fitting an ellipsoidal figure to the density (used by Kent). For the bulge we used terms up to $l = 8$, while for the disk terms as far as $l = 14$ needed to be retained in order to resolve the small angular extent of the disk at radii up to 4kpc.

Armed with this potential, we calculated the velocity dispersion and mean streaming velocity for the isotropic oblate rotator model (eqs. 5 and 4), in the same way as Kent did. These were evaluated on a dense grid in the meridional plane, after which the best-fit non-negative bilinear-segment DF that generates the disk and bulge density and isotropic-rotator streaming velocity was calculated. Parameters of the grid in the $(L, \mathcal{E}_{\mathrm{norm}})$-plane are given in Table 1. We assumed Kent's values of the bulge mass-to-light ratio, $\Upsilon_B = 1$, and adopted a disk mass-to-light ratio $\Upsilon_D = 0.87$ (corresponding to the disk surface density at the solar radius measured by Kuijken and Gilmore 1989b)[5]. The resulting DF is shown in Fig. 2. The rather patchy nature of this DF is reproduced with different choices for the grid parameters, and so the large stretches of phase space with zero density probably indicate that the mathematically exact inversion would give negative densities in places. The present DF represents (or would represent, in the limit of a very fine grid) the closest physically acceptable isotropic rotator model for Kent's potential and density. Large empty regions of phase space raise concerns about stability of the model; we shall not, however, address those here.

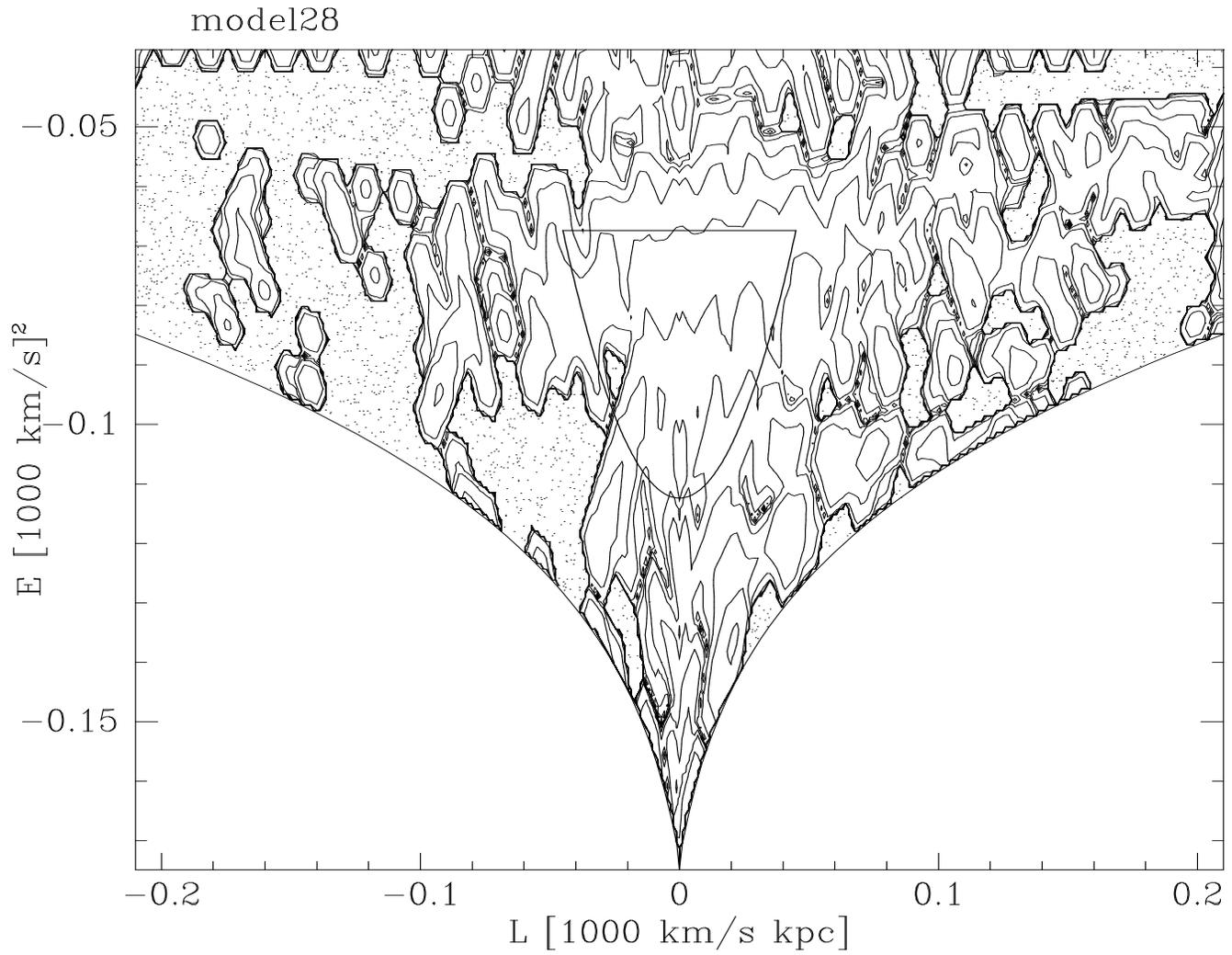

FIG. 2.–The best-fit isotropic oblate rotator DF for the central 1.2 kpc of the bulge. On adjacent contours, the DF differs by a factor of $10^{0.5}$. The thick parabola indicates the region of phase space accessible to stars with speeds below $300\,\mathrm{km\,s^{-1}}$ on the meridional plane in Baade's Window.

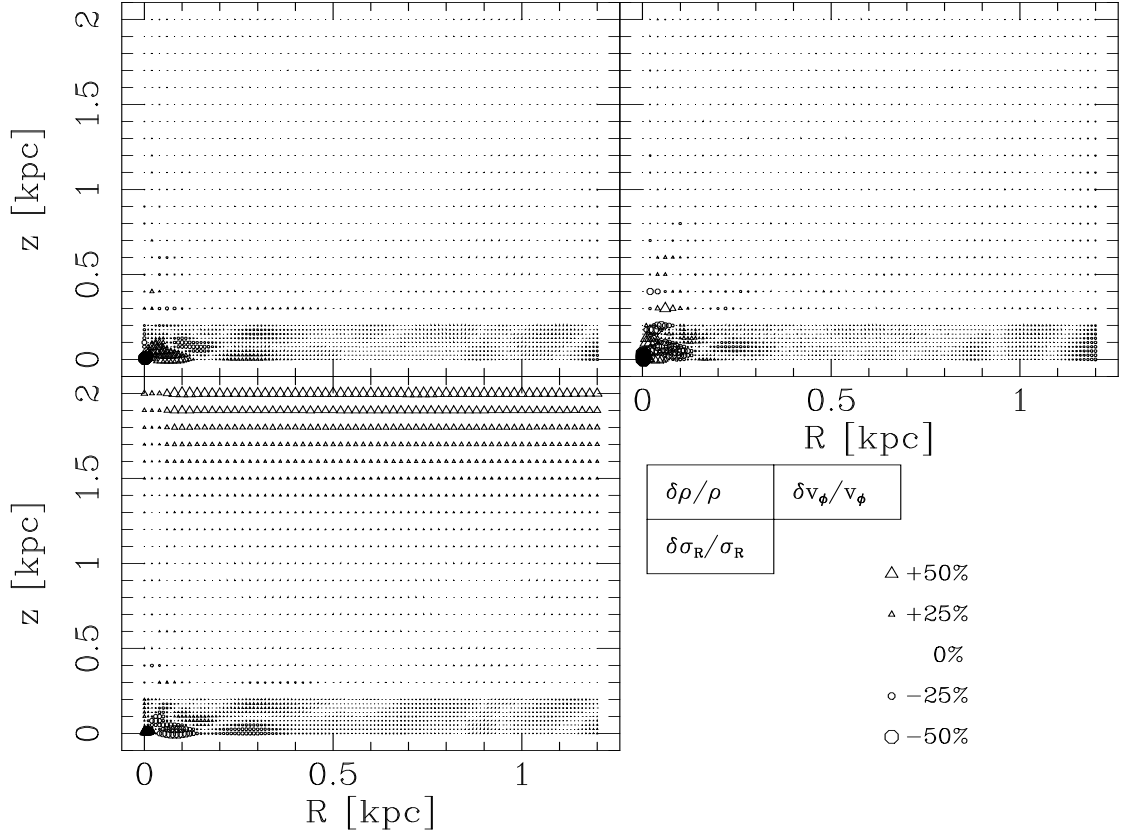

FIG. 3.–The deviations between the values of the density (top left), mean azimuthal velocity (top right) and radial velocity dispersion (bottom left) predicated by the oblate isotropic rotator model, and the values calculated from the best-fit bilinear-segment model. Circles indicate underestimates by the model, triangles overestimates. The size of the symbols is proportional to the discrepancy. The radial dispersion values were not fitted, but instead serve as a confirmation of the quality of the fit. The overestimate of the velocity dispersion at higher $z$ arises from the energy cutoff employed in the model fitting.

---

[5] Kent (1992) quotes $\Upsilon_D = 0.68$ as the value corresponding to this value for the local disk surface mass density: the difference arises from our sech $z$-profile replacing Kent's vertically exponential disk.

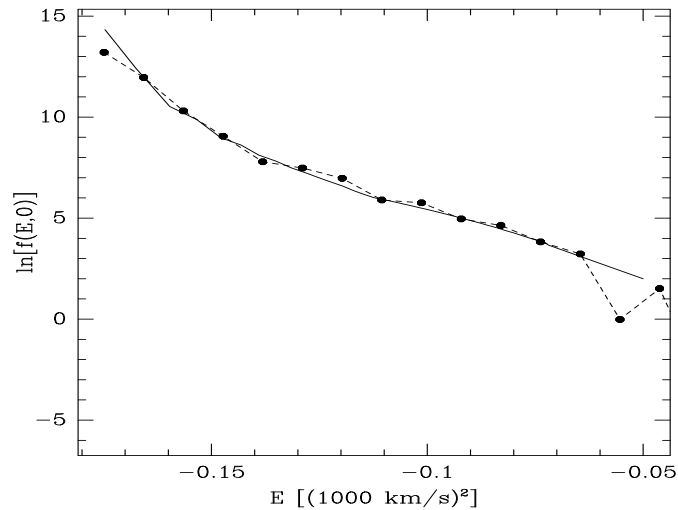

FIG. 4.–Comparison between the direct (Abel inversion) calculation of the DF at zero angular momentum, $f(E,0)$ (solid line), and the result of the bilinear-segment DF fit to the inner 1.2 kpc shown in figure 2 (dashed line with plotted points).

As can be seen in Fig. 3, which shows the deviation between the bilinear-segment DF and the values expected for the isotropic oblate rotator, a good fit to the density and isotropic rotator streaming velocity could be achieved in spite of the regions of zero phase density. The rms fractional error for the $\rho$- and $\rho\overline{v}_\phi$-values fitted was 10%, with most of that coming from the central 100pc where the resolution of the grid limited the ability of the model to follow the steep density gradient. Apart from the central regions, no large systematic residuals between the model and the fitted density and mean velocity were found. More importantly, the third panel in this figure shows that the velocity dispersions calculated from the model agree well with the isotropic rotator values. This was not directly enforced on the fit, and the good agreement therefore indicates that no part of the DF was missed by the finite extent of the grid or as a result of poor resolution. In the central parts of the bulge the finite resolution is again noticeble, however, and at large distances from the center the effect of the energy cutoff is evident as a systematically overestimated velocity dispersion.

The DF at zero angular momentum is related only to the density on the $z$-axis (see §3.2), and can be calculated directly (eq. 9). In Fig. 4 the result of this calculation is compared to the fit of the entire integral plane. The results agree well, except in the very center of the model where the $(E, L)$ grid lacks resolution.

Having obtained the DF, all kinematic quantities are specified and can be compared to observations. Thus, in Fig. 5 the radial velocity distribution of those stars with $R < 3$kpc in a small patch of sky towards Baade's Window (longitude 1°, latitude −4°) is shown, and Fig. 6 presents the distribution in proper motions for the same patch of sky. These model predictions can be compared with the data of Sharples et al. (1990) and of Spaenhauer et al. (1992), which Kent (1992) already showed have velocity dispersions similar to those predicted by the isotropic rotator model. It turns out that the model distributions are remarkably gaussian: in fact, a decent approximation to the part of phase space accessible to stars with speeds below $300 \mathrm{km\,s^{-1}}$ in the meridional plane at Baade's window (indicated by the parabola in Fig. 2) is $f \propto \exp[-E/(114 \mathrm{km\,s^{-1}})^2 + L/(78 \mathrm{km\,s^{-1}kpc})]$, which indeed produces a gaussian velocity distribution for all three components in the meridional plane. Since stars near the meridional plane dominate the density along the line of sight, the result is a rather gaussian projected velocity distribution.

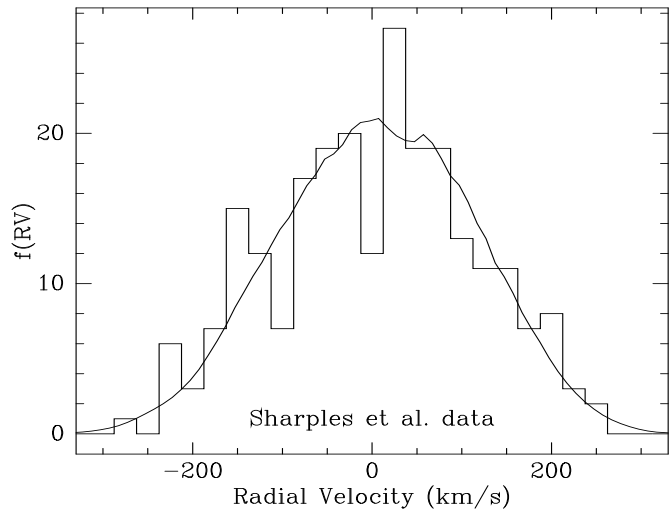

FIG. 5.–The observed heliocentric radial velocity distribution of M giants in Baade's Window (Sharples et al. 1990) and the distribution calculated from the bilinear-segment DF of figure 2. The agreement is remarkable.

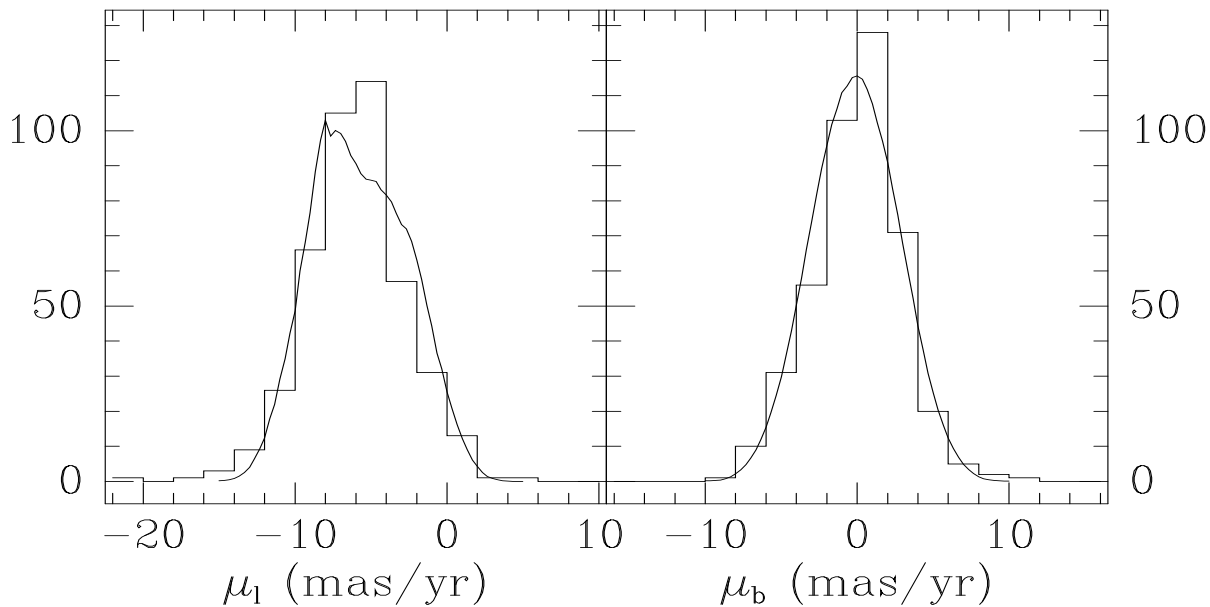

FIG. 6.–The observed proper-motion distribution in longitude, $\mu_l$, and latitude, $\mu_b$, of K giants in Baade's Window (Spaenhauer et al. 1992) and the distribution calculated from the bilinear-segment DF of figure 2. The zero point of the data (which is unknown) has been adjusted to agree with the model. The asymmetry in the distribution of observed latitudinal proper motion is not terribly significant, and cannot be reproduced in any steady-state model which does not invoke long-axis tube orbits in a triaxial bulge.

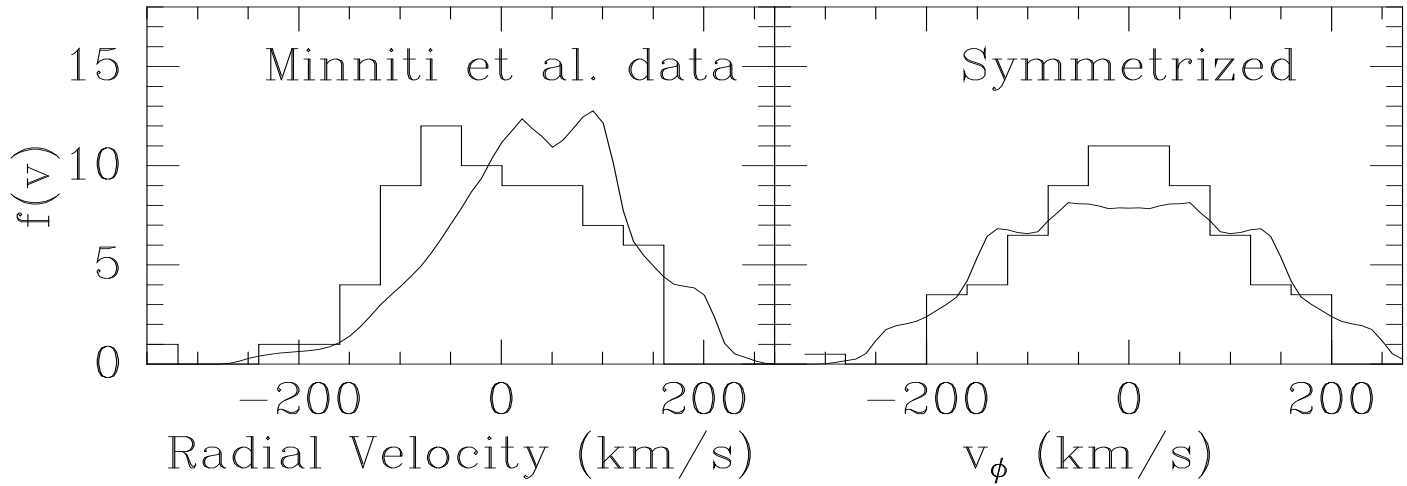

FIG. 7.–Left: The observed heliocentric radial velocity distribution (histogram) towards $l = 8$, $b = 7$, from Minniti et al. (1992), and the prediction of the model DF calculated here. Right: symmetrized distribution of *Galacto*centric radial velocity (see text for details).

In order to allow comparison with fields at slightly greater radii, a second model was calculated with a coarser grid: compared to the values in table 1, the angular momentum now extended to $0.77 \mathrm{km\,s^{-1}kpc}$, the value for a circular orbit with radius 4kpc, and the cutoff energy was raised to zero. The spacing of the $(R, z)$ values at which the density and mean velocity were fitted was also increased, by a factor of four. The number of bins remained unchanged. The predicted distribution of radial velocities for the resulting model, in the direction $l = 8$, $b = 7$, and the data of Minniti et al. (1992) are shown in figure 7. The mean LSR radial velocity and velocity dispersion of the stars are observed to be $16 \pm 10 \mathrm{km\,s^{-1}}$ and $85 \pm 7 \mathrm{km\,s^{-1}}$, (Minniti et al. quote a mean *galacto*centric velocity of $45 \mathrm{km\,s^{-1}}$) whereas the model DF calls for $43 \mathrm{km\,s^{-1}}$ and $90 \mathrm{km\,s^{-1}}$ (in excellent agreement with the prediction of Kent's model, as quoted by de Zeeuw 1993). In addition to having discrepant means, the radial velocity distributions of the data and model also have different shapes.

Overall, the model calls for faster rotation in the $l = 8$, $b = 7$ field than the data show. It is not clear how to interpret the discrepancy, given the number of modelling assumptions that went into the derivation of the DF: in particular, three-integral models will need to be constructed before it can be clear whether this is stellar-dynamical evidence for triaxiality or for a different gravitational potential than has been assumed here.

We can more easily assess whether a non-isotropic, but still two-integral, model may be able to explain the observations, by studying only the even part of the DF: as noted above, $f_+$ is the same for all two-integral DFs with the same density and potential. The radial velocity distribution of the even part of the DF is obtained simply by symmetrizing the radial velocity distribution in the galactocentric frame (taken to move at $220 \mathrm{km\,s^{-1}}$ with respect to the local standard of rest), and is shown together with the model in the right-hand panel of figure 7. It appears that the data in this field are not consistent with a two-integral, axisymmetric DF in the potential chosen.

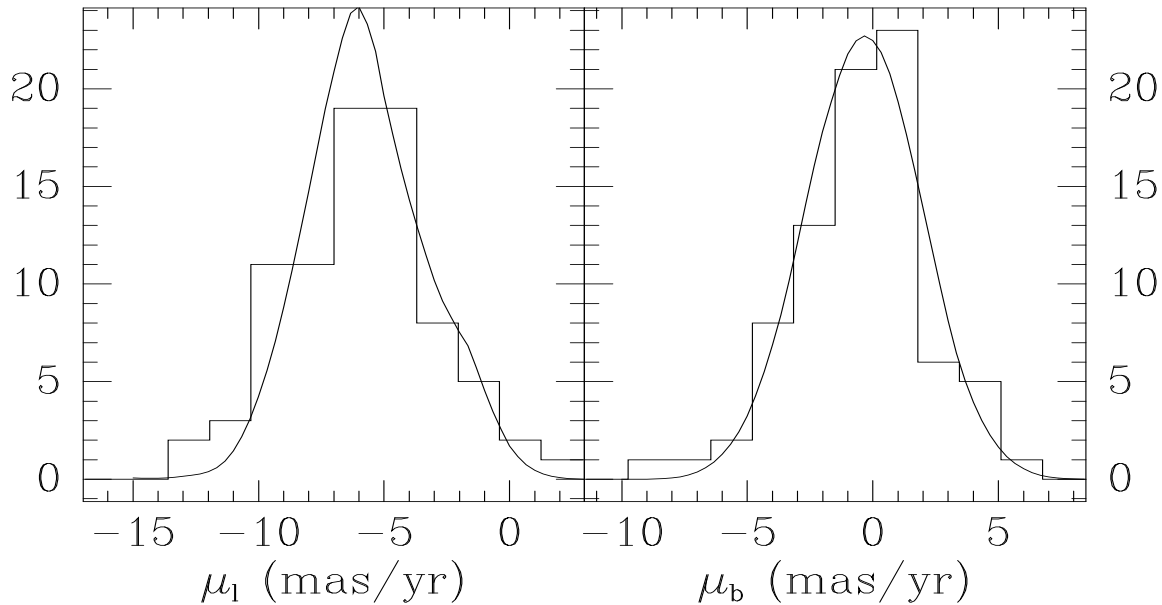

Fig. 8.–The model and observed proper-motion distributions towards $l = 8$, $b = 7$, taken from Minniti (1993). The proper motion zero points have been adjusted to give the best fit.

A similar, though smaller, discrepancy is evident in the proper motion distributions in the same field (figure 8). Note that, without good distance information, the distribution of $\mu_l$ is virtually independent of the odd part of the DF.

Minniti et al. also provide data for a field at $l = 12$, $b = 3$: however their correction for disk star contamination is serious in this direction, and hard to simulate. The agreement between the model and the observed velocities (not shown) is even poorer in this case.

## 5  DISCUSSION AND SUMMARY

Bilinear segments can be used to approximate any continuous function of two variables over a finite domain. Thus, spherical galaxies or clusters with DF $f(E, \mathbf{L}^2)$, axisymmetric galaxies with $f(E, L)$ (such as the bulge model discussed in this paper), or thin disks with $f(E, L)$, are candidates for this technique. An example of the last type is the disk DF for the galaxy NGC 7217, extracted from long-slit spectroscopy data by Merrifield & Kuijken (1994).

Our application to the bulge has shown that this technique makes it possible to calculate the distributions of observed quantities such as radial velocities or proper motions for the isotropic oblate rotator model. Thus, more detailed comparison can be made to observed data, and stronger conclusions drawn about the model, than is allowed by the use of the Jeans equations alone.

In principle, it would be possible to use these models to infer an $\mathcal{M}/\mathcal{L}$ value for the central part of the galaxy, for example by calculating a grid of models with different disk and bulge mass-to-light ratios, or by adding in varying amounts of dark matter to the total mass density. Comparison of the data in figures 5 and 6 with these models would then allow a statistical evaluation of the potential parameters. However, since the two-integral models, well though they fit, are still a small subset of possible models for the bulge kinematics, such a determination would still be subject to unknown systematic model uncertainties. Nevertheless, it appears that the adopted $2.4\mu$ mass-to-light ratios for the bulge and the disk (respectively, 1.0 and 0.87) are probably close to the truth.

Throughout this analysis, we have assumed that the bulge is axisymmetric. This is, in fact, unlikely to be the case: there is mounting evidence, from a variety of sources, that the bulge exhibits significant asymmetries between the positive- and negative-longitude sides (Blitz and Spergel 1991, Binney et al. 1991, Whitelock, Feast & Catchpole 1991, Weinberg 1992). The most natural way to understand the asymmetries is to posit that the bulge is a triaxial body, with long axis pointing to a Galactic azimuth of about 30°; projection effects then cause the nearer side of the bulge to have a higher scale height, and a lower surface brightness, than the far side. Kinematic studies of anomalies in the gas kinematics in the central few kpc independently point to quite similar (though not identical) triaxiality parameters for the bulge.

In the context of claims for triaxiality of the bulge, it is interesting that the data appear to show too little azimuthal motion, compared to the axisymmetric models: the favoured bar models place the sun nearest the bar's long axis, in which case, from our vantage point, the naive expectation would be of higher radial velocities in the bar model than in the non-barred one. This effect is offset to some extent by the errors made in fitting an axisymmetric potential: a bar seen edge-on and analysed assuming axisymmetry suggests a more centrally condensed potential than is really the case, and consequently axisymmetric models fitted to it predict circular velocities which are too high. A more detailed calculation is required in order to see which effect is the greater.

If the bulge is really triaxial, it is certainly remarkable that the radial velocity and the proper motion data are fit excellently with an oblate isotropic rotator, arguably the simplest axisymmetric kinematic bulge model possible. Nevertheless, this may be no more than a coincidence, given that in at least one other field ($l = 8$, $b = 7$) the fit is not nearly so good—but the sample is relatively small. More data in a number of fields will be required to settle this issue definitely. But even if the oblate isotropic rotator model is ruled out, this leaves many more axisymmetric models as possibilities. It will be interesting to see how firm stellar-dynamical evidence for non-axisymmetry can be. It will be especially worthwhile to find signatures for non-axisymmetry which do not rely on perspective, so that they can be applied to external galaxies.

This work was supported by a Hubble Fellowship through grant HF-1020.01-91A awarded by the Space Telescope Science Institute (which is operated by the Association of Universities for Research in Astronomy, Inc., for NASA under contract NAS5-26555).

## APPENDIX: KINEMATICS ON THE MINOR AXIS

In this appendix, we show that on the minor axis of an axisymmetric stellar system, as well as in two-integral axisymmetric models in general, the vertical kinematics are closely analogous to those of a plane-parallel system.

The collisionless Boltzman equation in cylindrical polar coordinates in a gravitational field with forces $\mathcal{K}_R$ and $\mathcal{K}_z$ reads (e.g., Binney and Tremaine 1987)

$$\frac{\partial f}{\partial v_R}\left(\mathcal{K}_R + \frac{v_\phi^2}{R}\right) + \frac{\partial f}{\partial v_\phi}\left(-\frac{v_R v_\phi}{R}\right) + \frac{\partial f}{\partial v_z}\mathcal{K}_z + \frac{\partial f}{\partial R}v_R + \frac{\partial f}{\partial z}v_z = 0. \tag{12}$$

We set $F_z(R, z, v_z) = \int f\,dv_R dv_\phi$. Integrating eq. (12) over $v_R$ and $v_\phi$, and assuming that the phase space density tends to zero at infinite velocity, we obtain

$$\frac{1}{R}\frac{\partial}{\partial R}R\int f v_R\,dv_R dv_\phi + \frac{\partial F_z}{\partial v_z}\mathcal{K}_z + \frac{\partial F_z}{\partial z}v_z = 0. \tag{13}$$

The cartesian version of this equation is

$$\int \left(v_x\frac{\partial f}{\partial x} + v_y\frac{\partial f}{\partial y}\right)dv_x dv_y + \frac{\partial F_z}{\partial v_z}\mathcal{K}_z + \frac{\partial F_z}{\partial z}v_z = 0. \tag{14}$$

The quantity $\int f v_R\,dv_R dv_\phi$ depends on $v_z$, and is proportional to the mean $R$-velocity for a given $v_z$. It is not generally zero (since the third integral usually introduces a non-zero correlation between the $R$- and $z$-velocity components, known as the 'velocity ellipsoid tilt'), but in two important cases it is: on the minor axis of any non-singular axisymmetric system (this is best illustrated with eq. 14 upon setting the partial $x$- and $y$-derivatives of $f$ to zero), and in the case of a two-integral DF $f(E, L)$ such as has been considered in this paper. In these special circumstances $F_z$ satisfies the one-dimensional collisionless Boltzman equation

$$\mathcal{K}_z\frac{\partial F_z}{\partial v_z} + v_z\frac{\partial F_z}{\partial z} = 0, \tag{15}$$

just as is the case for a plane-parallel stellar slab. This equation can be integrated easily: the general solution is

$$F_z = F_z(R, E_z) \qquad \text{where} \quad E_z = \Psi(R, z) + \tfrac{1}{2}v_z^2. \tag{16}$$

Another way to prove this result in the case of the two-integral models follows from the fact that $F_z = \int dv_R dv_\phi f(E, L)$ can be written as

$$\int_{-\infty}^{\infty} dv_\phi \int_0^{\infty} dv_R\, f(\tfrac{1}{2}v_R^2 + \tfrac{1}{2}v_\phi^2 + E_z, Rv_\phi) \equiv F_z(R, E_z). \tag{17}$$

At every radius, $F_z(R, z, v_z)$ can therefore be obtained by finding the $E_z$ dependence that generates the correct vertical density profile,

$$\rho(R, z) = 2\int_\Psi(R, z)\frac{dE_z}{\sqrt{2[E_z - \Psi(R, z)]}}F_z(R, E_z). \tag{18}$$

This solution can be found with a single Abel inversion (Kuijken and Gilmore 1989a). In the case of two-integral models, the $v_R$-distribution follows immediately, since it is identical to the distribution of $v_z$.

The minor axis of an axisymmetric system is particularly interesting, since there the result (16) holds independent of the two-integral assumption. A measurement of the proper motion distribution of stars on the minor axis of the bulge would help greatly to define the gravitational potential on that axis.